\documentclass[11pt]{article}

\usepackage{epsfig}

\newcommand{\gfp}{GF($p$)}

\begin{document}

\title{Quantum Addition Circuits and Unbounded Fan-Out}

\author{Yasuhiro Takahashi\footnote{NTT Communication Science
Laboratories, NTT Corporation}
\hspace{1cm}
Seiichiro Tani$^*$\footnote{Quantum Computation and Information Project,
ERATO-SORST, JST}
\hspace{1cm}
Noboru Kunihiro\footnote{Graduate School of Frontier Sciences, The
University of Tokyo}}

\date{}

\maketitle

\begin{abstract}
We first show how to construct an $O(n)$-depth $O(n)$-size quantum
circuit for addition of two $n$-bit binary numbers with no ancillary
qubits. The exact size is $7n-6$, which is smaller than that of any
other quantum circuit ever constructed for addition with no ancillary
qubits. Using the circuit, we then propose a method for constructing an
$O(d(n))$-depth $O(n)$-size quantum circuit for addition with
$O(n/d(n))$ ancillary qubits for any $d(n) = \Omega(\log n)$. If we are
allowed to use unbounded fan-out gates with length $O(n^{\varepsilon})$
for an arbitrary small positive constant $\varepsilon$, we can modify
the method and construct an $O(e(n))$-depth $O(n)$-size circuit with
$o(n)$ ancillary qubits for any $e(n) = \Omega(\log^* n)$. In
particular, these methods yield efficient circuits with depth $O(\log
n)$ and with depth $O(\log^* n)$, respectively. We apply our circuits to
constructing efficient quantum circuits for Shor's discrete logarithm
algorithm.
\end{abstract}

\section{Introduction}

Since Shor's discovery of quantum algorithms for factoring and discrete
logarithm problems \cite{Shor}, many studies have investigated ways of
constructing quantum circuits for the algorithms
\cite{Vedral,Zalka,Beauregard,Proos,Fowler,Takahashiorder}. The
resulting circuits are important not only for implementing the
algorithms on a quantum computer but also for understanding the
computational power of small quantum circuits. These studies have shown
that addition of two binary numbers is a key operation for constructing
quantum circuits for Shor's algorithms.

We consider the problem of constructing quantum circuits for addition of
two binary numbers with better complexity. The complexity measures of a
quantum circuit are its size and depth, and the number of qubits in
it. Roughly speaking, the size and depth correspond to computation time,
while the number of qubits corresponds to the size of memory. We regard
the number of qubits as a primary consideration since it seems difficult
to realize a quantum computer with many qubits. It is not obvious
whether the number of qubits in a quantum circuit for addition can be
decreased by using efficient classical ones, though the size or depth
can be decreased simply by using them.

An unbounded fan-out gate on $n+1$ qubits copies a classical source bit
into $n$ copies. In particular, the gate on two qubits is a CNOT
gate. If unbounded fan-out gates are available, sublogarithmic-depth
quantum circuits for various operations can be constructed
\cite{Green,Hoyer}. This is because the gate performs the copy operation
on an unbounded number of qubits in a constant time. However, it seems
difficult to realize such a gate practically. Thus, it is important to
minimize the number of target qubits of the gate in a circuit without
increasing the complexity of the circuit. When we use unbounded fan-out
gates, we consider the complexity measures (size, depth, and the number
of qubits) for the number of target qubits of the gate. We call the
number of target qubits the length of an unbounded fan-out gate.

There have been many studies of efficient quantum circuits for addition
of two $n$-bit binary numbers. These circuits can be classified
according to depth complexity. Draper's and Takahashi et al.'s circuits
have depth $O(n)$ and use no ancillary qubits
\cite{Draper,Takahashiaddition}. Takahashi et al.'s is more efficient
than Draper's since the sizes of Takahashi et al.'s and Draper's are
$O(n)$ and $O(n^2)$, respectively. Draper et al.'s and Takahashi et
al.'s circuits have depth $O(\log n)$ \cite{Svore,Takahashiadd2}. Draper
et al.'s uses $O(n)$ ancillary qubits and its size is $O(n)$. Takahashi
et al.\ decreased the number of ancillary qubits to $O(n/\log n)$ without
increasing the size asymptotically. H\o yer et al.\ showed that, if
unbounded fan-out gates with length $O(n)$ are available, an $O(\log^*
n)$-depth circuit can be constructed \cite{Hoyer}. They have not
analyzed the number of ancillary qubits or size.

In this paper, we first show how to construct an $O(n)$-depth
$O(n)$-size quantum circuit for addition with no ancillary qubits. The
circuit is based on the ripple-carry approach. The exact size is $7n-6$,
which is smaller than that of any other quantum circuit ever constructed
for addition with no ancillary qubits. Moreover, the circuit is more
implementable than the previous circuits with no ancillary qubits in the
sense that the circuit can be used directly on a linear nearest neighbor
architecture \cite{Fowler}, i.e., on a unidimensional array of qubits
with nearest neighbor interactions only. By combining the circuit with
the carry-lookahead approach, we then propose a method for constructing
an $O(d(n))$-depth $O(n)$-size quantum circuit for addition with
$O(n/d(n))$ ancillary qubits for any $d(n)=\Omega(\log n)$. The method
is a generalized and simplified version of Takahashi et al.'s method for
constructing a logarithmic-depth circuit with a small number of qubits
\cite{Takahashiadd2}. In particular, for $d(n)=\log n$, our method
yields an $O(\log n)$-depth $O(n)$-size circuit with $O(n/\log n)$
ancillary qubits. The number of ancillary qubits is exactly the same as
that in Takahashi et al.'s circuit and the size is less than half that
of Takahashi et al.'s.

If we are allowed to use unbounded fan-out gates with length
$O(n^{\varepsilon})$ for an arbitrary small positive constant
$\varepsilon$, we can modify our method and construct an $O(e(n))$-depth
$O(n)$-size circuit with $O(n\log^{**}n/e(n))$ ancillary qubits for any
$e(n)=\Omega(\log^* n)$, where $\log^{**}n$ is a slowly-growing function
satisfying $\log^{**}n=o(\log^* n)$. The main point of this modification
is to decrease the depth of the carry-lookahead part of our method by
using a quantum version of Chandra et al.'s constant-depth classical
circuit for addition with unbounded fan-in and fan-out gates
\cite{Chandra}. To construct the quantum version, we require a quantum
gate corresponding to an unbounded fan-in gate. We use H\o yer et al.'s
small-depth quantum circuit for a generalized Toffoli operation with
unbounded fan-out gates \cite{Hoyer} as the gate. In particular, for
$e(n)=\log^* n$, the modified method yields an $O(\log^* n)$-depth
$O(n)$-size circuit with $o(n)$ ancillary qubits. Though H\o yer et al.\
have constructed an $O(\log^* n)$-depth circuit for addition as
mentioned above, our construction shows that the number of ancillary
qubits, size, and the length of an unbounded fan-out gate can be small
simultaneously.

This construction also shows that unbounded fan-out gates with a small
length are sufficient to construct a sublogarithmic-depth circuit. For
example, if we are allowed to use unbounded fan-out gates with length
$O(\log n)$, we can construct an $O(\log n/\log\log n)$-depth
$O(n)$-size circuit with $o(n)$ ancillary qubits. Such a
sublogarithmic-depth circuit cannot be constructed by using a quantum
circuit only with gates on a bounded number of qubits \cite{Fang} or by
using a classical circuit only with bounded fan-in and unbounded fan-out
gates \cite{Pippenger}.

Using our circuits for addition, we construct efficient quantum circuits
for Shor's discrete logarithm algorithm for elliptic curves over the
prime field \gfp. This is done by simply using our addition circuits in
Proos et al.'s circuit for Shor's discrete logarithm algorithm
\cite{Proos}. Since Proos et al.'s circuit uses $n$ ancillary qubits
during addition, the use of our circuit with no ancillary qubits
decreases the $n$ ancillary qubits without increasing the original depth
or size asymptotically, where $n$ is the length of the binary
representation for $p$. Moreover, we decrease the depth asymptotically
by adding $o(n)$ ancillary qubits. Proos et al.'s circuit with our
addition circuits is more efficient than with the previous ones
described above.

In contrast to the previous methods for constructing efficient quantum
circuits for addition
\cite{Draper,Takahashiaddition,Svore,Takahashiadd2,Hoyer}, our method is
general in the sense that it can yield various types of efficient
quantum circuits for addition. The generality allows us to construct
quantum circuits appropriate for various situations we will have to
consider practically. For example, if we want to save the number of
qubits, we can obtain a qubit-efficient circuit by setting $d(n)=n$ in
our method. We can decrease the depth by setting $d(n)=\log
n$. Moreover, we can choose an ``intermediate'' circuit by setting
$d(n)=\sqrt{n}$.

\section{Circuit with Depth $O(n)$}

\subsection{Ripple-Carry Approach}

We use the standard notation for quantum states and the standard
diagrams for quantum circuits \cite{Nielsen}. As mentioned earlier, the
measures of the complexity of a quantum circuit are the number of qubits
and its size and depth. The meaning of the number of qubits is
obvious. The size of a circuit is defined as the total number of
elementary gates in it. The elementary gates are one-qubit unitary
gates, CNOT gates, controlled-$R_t$ gates, and Toffoli gates, where
$R_t|x\rangle = e^{2\pi i x/2^t}|x\rangle$ for $t \geq 1$ and $x \in
\{0,1\}$. In Section 4, we use the gate for an unbounded fan-out
operation $F_t$ as an elementary gate, where $F_t$ (on $t+1$ qubits) is
defined as
$$F_t\left(|y\rangle\bigotimes_{i=0}^{t-1}|x_i\rangle\right)
=|y\rangle\bigotimes_{i=0}^{t-1}|x_i\oplus y\rangle$$
for $y,x_i \in \{0,1\}$. The symbol $\oplus$ denotes addition modulo
2. The depth of a circuit is defined as follows. Input qubits are
considered to have depth 0. For each gate $G$, the depth of $G$ is equal
to 1 plus the maximal depth of a gate on which $G$ depends. The depth of
a circuit is equal to the maximal depth of a gate in it. Intuitively,
the depth is the number of layers in the circuit, where a layer consists
of gates that can be performed simultaneously. A quantum circuit can use
ancillary qubits, which start and end in the state $|0\rangle$. We
usually count the number of ancillary qubits instead of the number of
all qubits used in the circuit.

We consider the problem of constructing quantum circuits for the
operation ADD$_n$ defined as
$$\left(\bigotimes_{i=0}^{n-1}|b_i\rangle|a_i\rangle\right)|z\rangle \to
\left(\bigotimes_{i=0}^{n-1}|s_i\rangle|a_i\rangle\right)|z \oplus
s_n\rangle,$$
where $a_{n-1}\cdots a_0$ and $b_{n-1}\cdots b_0$ are the input binary
numbers, $z \in \{0,1\}$, and $s_n\cdots s_0$ is the sum of the input
binary numbers. Our linear-depth circuit and most of the previous ones
with a small number of qubits are based on the ripple-carry approach. To
explain the approach, we define the carry bit $c_i$ $(0 \leq i \leq n)$
as follows:
$$ c_i = \left\{
\begin{array}{ll}
0  & \mbox{$i=0$,}\\
{\rm MAJ}(a_{i-1},b_{i-1},c_{i-1}) &
 \mbox{$1\leq i \leq n$,}
\end{array}
\right.$$
where MAJ is the majority function for three bits defined as ${\rm
MAJ}(a,b,c)=ab \oplus bc \oplus ca$. In the ripple-carry approach, the
first step is to compute the carry bit $c_1$ by using $a_0$ and $b_0$
and $c_0$. Then, $c_2$ is computed by using $a_1$ and $b_1$ and
$c_1$. This procedure is repeated until all carry bits are
computed. After that, $s_i$ $(0 \leq i \leq n)$ is computed by the
relationship
$$ s_i = \left\{
\begin{array}{ll}
a_i \oplus b_i \oplus c_i  & \mbox{$0 \leq i \leq n-1$,}\\
c_n & \mbox{$i=n$.}
\end{array}
\right.$$

\begin{figure}
 \vspace{-2.1cm}
\centering
\epsfig{file=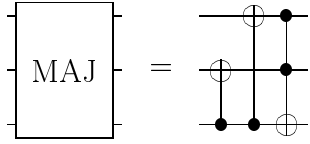,scale=.5}
\vspace{-6.3cm}
\caption{The MAJ gate.}
\label{figure1}
\end{figure}

When the ripple-carry approach is used, the key issue for constructing a
quantum circuit with a small number of qubits is how to store carry
bits. Cuccaro et al.'s circuits, which are based on the approach, use
one ancillary qubit to store $c_0=0$ \cite{Cuccaro}. The carry bit $c_i$
is stored in the qubit initially storing $a_{i-1}$ for $1 \leq i \leq
n$. To do this, they defined the gate for MAJ depicted in
Fig. \ref{figure1}, which is the main component of their circuits. The
gate maps $|c_i\rangle |b_i\rangle |a_i\rangle$ to $|c_i \oplus
a_i\rangle |b_i \oplus a_i\rangle |c_{i+1}\rangle$. Takahashi et al.'s
circuit, which is also based on the ripple-carry approach, uses no
ancillary qubits \cite{Takahashiaddition}. All the carry bits are stored
in the qubit initially storing $z$. The main component of their circuit
is also the MAJ gate. They use the property that the gate maps $|z
\oplus b_i\rangle |z \oplus a_i\rangle |z \oplus c_i\rangle$ to $|b_i
\oplus c_i\rangle |a_i \oplus c_i\rangle |z \oplus c_{i+1}\rangle$.

\subsection{Our Circuit}

We store the carry bit $c_i$ in the qubit initially storing $a_i$ for $0
\leq i \leq n-1$ and store the high-order bit $c_n$ in the qubit
initially storing $z$. This would be difficult to do if we use the MAJ
gate directly. Our idea is to divide the MAJ gate into two parts. The
first part consists of two CNOT gates and the second one consists of one
Toffoli gate. It is easy to verify that a Toffoli gate maps $|b_i \oplus
a_i\rangle |a_i \oplus c_i\rangle |a_{i+1} \oplus a_i\rangle$ to $|b_i
\oplus a_i\rangle |a_i \oplus c_i\rangle |a_{i+1} \oplus c_{i+1}\rangle$
for $1 \leq i \leq n-1$, where we consider $a_n$ as $z$. Thus, using
CNOT gates (the first parts of the MAJ gate) and a Toffoli gate, we
first prepare the state
$$|b_1 \oplus a_1\rangle |a_1 \oplus c_1\rangle
\left(\bigotimes_{i=2}^{n-1}|b_i \oplus a_i\rangle |a_i \oplus
a_{i-1}\rangle\right)|z \oplus a_{n-1}\rangle.$$
By applying Toffoli gates (the second parts of the MAJ gate), we can
compute $c_i$ and store it in the qubit initially storing $a_i$. The
final Toffoli gate computes $c_n$ and stores it in the qubit initially
storing $z$. The detailed construction is described below.

\begin{figure}
\vspace{-1.6cm}
\centering
\epsfig{file=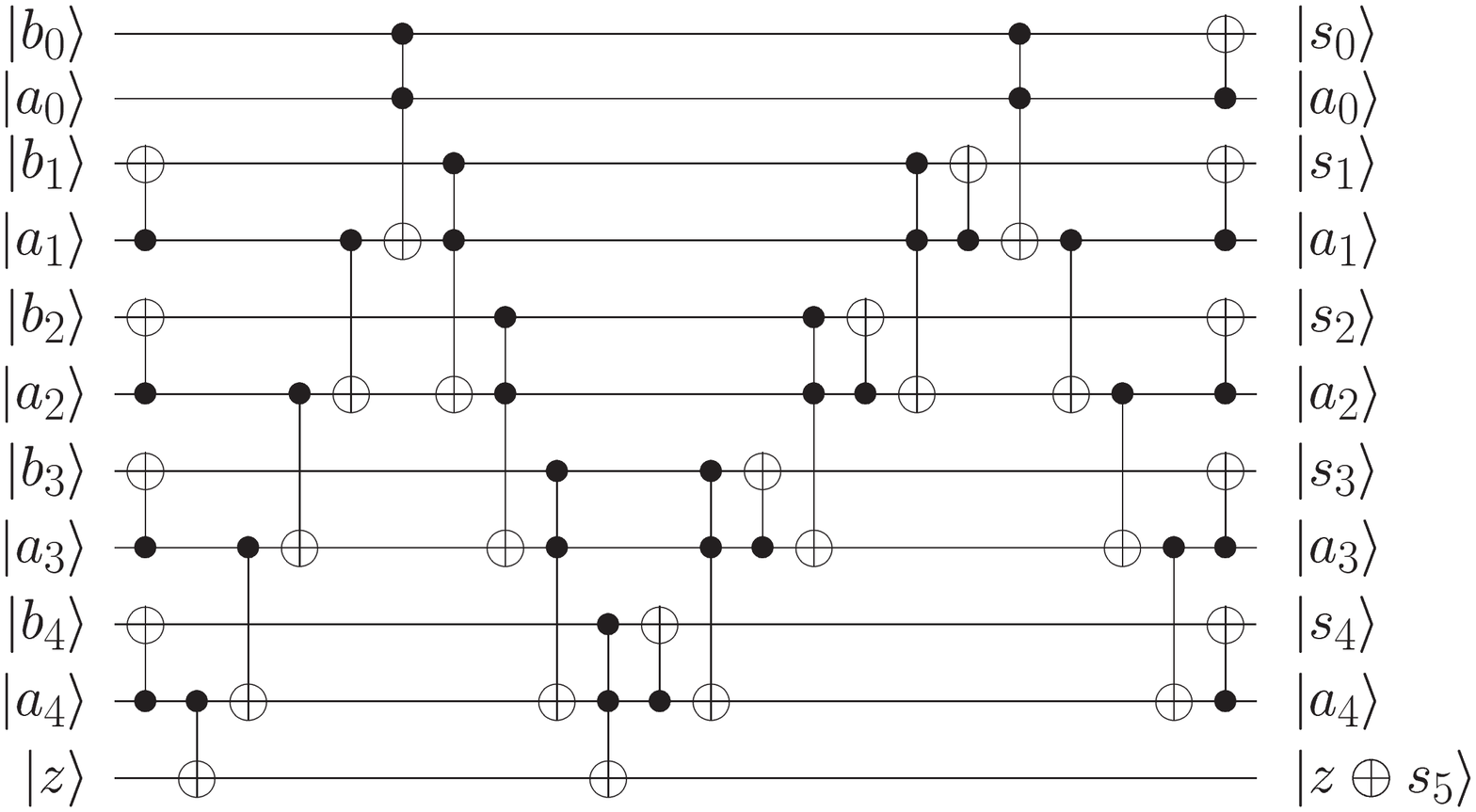,scale=0.37}
\vspace{-1.3cm}
\caption{The circuit for ADD$_5$.}
\label{figure2}
\end{figure}

Let $A_i$ and $B_i$ denote the memory locations initially storing $a_i$
and $b_i$, respectively, for $0 \leq i \leq n-1$. Let $A_n$ be the
memory location initially storing $z$. Location $A_i$ $(0 \leq i \leq
n-1)$ will store $a_i$, $B_i$ $(0 \leq i \leq n-1)$ will store $s_i$,
and $A_n$ will store $z\oplus s_n$ at the end of the computation. Our
circuit is constructed in the following six steps.
\begin{enumerate}
\item For $i = 1,\ldots,n-1$:

Apply a CNOT gate to a pair of memory locations $B_i$ and $A_i$ where
      $A_i$ is used for the control qubit.

\item For $i = n-1,\ldots,1$:

Apply a CNOT gate to a pair of memory locations $A_i$ and $A_{i+1}$
      where $A_i$ is used for the control qubit.

\item For $i = 0,\ldots,n-1$:

Apply a Toffoli gate to a tuple of memory locations $B_i$, $A_i$ and
      $A_{i+1}$, where $B_i$ and $A_i$ are used for the control qubit.

\item For $i = n-1,\ldots,1$:

Apply a CNOT gate to a pair of memory locations $B_i$ and $A_i$ where
      $A_i$ is used for the control qubit. Then, apply a Toffoli gate to
      a tuple of memory locations $B_{i-1}$, $A_{i-1}$ and $A_i$, where
      $B_{i-1}$ and $A_{i-1}$ are used for the control qubit.

\item For $i = 1,\ldots,n-2$:

Apply a CNOT gate to a pair of memory locations $A_i$ and $A_{i+1}$
      where $A_i$ is used for the control qubit.

\item For $i = 0,\ldots,n-1$:

Apply a CNOT gate to a pair of memory locations $B_i$ and $A_i$ where
      $A_i$ is used for the control qubit.
\end{enumerate}
The circuit for ADD$_5$ is depicted in Fig. \ref{figure2}.

We describe the changes of the input state of ADD$_n$ to show that the
circuit works correctly. In Step 1, the input state is transformed into
$$|b_0\rangle|a_0\rangle \left(\bigotimes_{i=1}^{n-1}|b_i \oplus a_i\rangle
|a_i\rangle\right)|z\rangle.$$
In Step 2, the state is transformed into
$$|b_0\rangle|a_0\rangle |b_1 \oplus a_1\rangle
|a_1\rangle\left(\bigotimes_{i=2}^{n-1}|b_i \oplus a_i\rangle |a_i
\oplus a_{i-1}\rangle\right)|z \oplus a_{n-1}\rangle.$$
The first Toffoli gate in Step 3 transforms the state into
$$|b_0\rangle|a_0\rangle|b_1 \oplus a_1\rangle |a_1 \oplus c_1
\rangle \left(\bigotimes_{i=2}^{n-1}|b_i \oplus a_i\rangle |a_i \oplus
a_{i-1}\rangle\right)|z \oplus a_{n-1}\rangle.$$
This is repeated by using a Toffoli gate. The state after Step 3 is
$$|b_0\rangle|a_0\rangle \left(\bigotimes_{i=1}^{n-1}|b_i \oplus
a_i\rangle |a_i \oplus c_i\rangle\right)|z \oplus s_n\rangle.$$
In Step 4, the state is transformed into
$$|b_0\rangle|a_0\rangle |b_1 \oplus c_1\rangle
|a_1\rangle\left(\bigotimes_{i=2}^{n-1}|b_i \oplus c_i\rangle |a_i
\oplus a_{i-1}\rangle\right)|z \oplus s_n\rangle.$$
In Step 5, the state is transformed into
$$|b_0\rangle|a_0\rangle \left(\bigotimes_{i=1}^{n-1}|b_i \oplus
c_i\rangle |a_i\rangle\right)|z \oplus s_n\rangle.$$
Since $s_i = a_i \oplus b_i \oplus c_i$ for $0 \leq i \leq n-1$, the
final step gives us the desired output state.

\subsection{Complexity Analysis}

From the construction, it is obvious that our circuit uses no ancillary
qubits. We compute the depth and size of the circuit for $n \geq 3$
precisely. In Step 1, the number of CNOT gates is $n-1$ and these gates
can be performed simultaneously. Thus, the depth and size of Step 1 are
1 and $n-1$, respectively. In Step 2, the number of CNOT gates is $n-1$
and thus the depth and size of Step 2 are $n-1$. In Step 3, the number
of Toffoli gates is $n$ and thus the depth and size of Step 3 are
$n$. In Step 4, the number of CNOT gates is $n-1$ and the number of
Toffoli gates is $n-1$. Thus, the depth and size of Step 4 are
$2n-2$. In Step 5, the number of CNOT gates is $n-2$ and thus the depth
and size of Step 5 are $n-2$. In Step 6, the number of CNOT gates is
$n$ and these gates can be performed simultaneously. Thus, the depth and
size of Step 6 are 1 and $n$, respectively. Thus, the depth and size of
the whole circuit are $5n-3$ and $7n-6$, respectively. The numbers of
CNOT and Toffoli gates are $5n-5$ and $2n-1$, respectively.

As discussed in \cite{Fowler}, many proposed quantum computer
architectures deal with a unidimensional array of qubits with nearest
neighbor interactions only. Thus, it is important for a circuit to work
on such a linear nearest neighbor (LNN) architecture. When the input and
output binary numbers are arranged on an LNN architecture in an
interleaved manner (as in Fig. \ref{figure2}), our circuit can be used
directly on an LNN architecture in the sense that the circuit can be
transformed into one on an LNN architecture without increasing the size
or depth asymptotically.

A comparison of our circuit and the previous ones with a small number of
qubits is summarized in Table \ref{Table1}. The symbol ``$\surd$'' in
the LNN column means that the circuit can be used directly on an LNN
architecture in the sense described above. The symbol ``---'' means that
we do not know whether this is the case for the circuit. The size of our
circuit is less than that of any other quantum circuit ever constructed
for ADD$_n$ with no ancillary qubits. When we regard the number of
qubits as a primary consideration, our circuit is more efficient than
the previous circuits in Table \ref{Table1}.

Though there exists a size-efficient or depth-efficient circuit with one
ancillary qubit \cite{Cuccaro}, it is worth noting that the difference
between the total number of ancillary qubits used by parallel
applications of our circuit (as in the next section) and that of the
previous circuit with one ancillary qubit depends on the number of
circuits applied in parallel and may become large. Moreover, since
Toffoli gates are on three qubits and thus may be harder to implement
than the other gates (on a smaller number of qubits), it is worth noting
that the number of Toffoli gates in our circuit is $2n-1$, which is less
than or equal to those of the previous circuits in Table \ref{Table1}
(excluding Draper's $O(n^2)$-size circuit).

\begin{table}
\caption{Comparison of Our Circuit and Previous Circuits}
\centering
\begin{tabular}{|c|c|c|c|c|c|} \hline
Circuit & Ancilla & Size & Toffoli & Depth &LNN \\ \hline
Cuccaro et al. \cite{Cuccaro} & 1 & $6n+1$ & $2n$ & $6n+1$ & $\surd$ \\
Cuccaro et al. \cite{Cuccaro} & 1 & $9n-8$ & $2n-1$ & $2n+4$ & $\surd$
 \\
Draper \cite{Draper} & 0 & $1.5n^2 + 4.5n+2$ & 0 & $5n+3$ & --- \\
Takahashi et al. \cite{Takahashiaddition} & 0 & $10n-9$ & $4n-5$ &
 $8n-7$ & --- \\
Our Circuit & 0 & $7n-6$ & $2n-1$ & $5n-3$ & $\surd$ \\ \hline
\end{tabular}
\label{Table1}
\end{table}

\section{General Method}

\subsection{Combination Method}

The ripple-carry approach decreases the number of ancillary qubits but
requires large depth. The carry-lookahead approach decreases the depth
but requires many qubits \cite{Svore}. Our method is based on the
combination of these methods and is a generalized and simplified version
of Takahashi et al.'s method for constructing a logarithmic-depth
circuit with a small number of qubits \cite{Takahashiadd2}. In this
section, we review the previous method. The carry-lookahead approach is
described by using two bits $p[i,j]$ $(1\leq i < j \leq n)$ and $g[i,j]$
$(0\leq i < j \leq n)$ \cite{Svore}. The bit $p[i,j]$ is 1 if a carry
bit is propagated from bit position $i$ to bit position $j$, and
$g[i,j]$ is 1 if a carry bit is generated between bit positions $i$ and
$j$. The $p[i,j]$ and $g[i,j]$ are computed by the following relations:
\begin{itemize}
\item For any $i$ such that $1\leq i \leq n-1$, $p[i,i+1]=a_i\oplus
      b_i$.

\item For any $i,j$ such that $1\leq i < i+1 < j \leq n$,
      $p[i,j]=p[i,t]p[t,j]$ for any $t$ satisfying $i<t<j$.
\item For any $i$ such that $0\leq i \leq n-1$, $g[i,i+1]=a_ib_i$.

\item For any $i,j$ such that $0\leq i < i+1 < j \leq n$,
      $g[i,j]=g[i,t]p[t,j]\oplus g[t,j]$ for any $t$ satisfying
      $i<t<j$.
\end{itemize}
It holds that $g[0,j] = c_j$ for all $1 \leq j \leq n$.

Draper et al.'s quantum carry-lookahead adder first computes $p[i,i+1]$
$(1\leq i \leq  n-1)$ and $g[i,i+1]$ $(0\leq i \leq n-1)$. Then, it
computes $g[0,i]$ $(1 \leq i \leq n)$ by successively doubling the sizes
of the intervals under consideration. Lastly, it computes $s_i$ $(0 \leq
i \leq n)$, where $s_0=p[0,1]$, $s_i=p[i,i+1]\oplus g[0,i]$ $(1\leq i
\leq n-1)$, and $s_n=g[0,n]$. The key circuit is the one for the second
step. We call this circuit the CARRY$_1$ gate. In general, the CARRY$_l$
gate is a circuit for the operation
$$\bigotimes_{i=1}^{\lfloor n/2^{l-1}\rfloor -1}|p_{l-1}[i]\rangle
\bigotimes_{j=0}^{\lfloor n/2^{l-1}\rfloor -1}|g_{l-1}[j]\rangle
\to \bigotimes_{i=1}^{\lfloor n/2^{l-1}\rfloor -1}|p_{l-1}[i]\rangle
 \bigotimes_{j=0}^{\lfloor n/2^{l-1}\rfloor
 -1}|g[0,2^{l-1}(j+1)]\rangle,$$
where $1 \leq l \leq \lfloor \log n \rfloor -1$,
$p_{l-1}[i]=p[2^{l-1}i,2^{l-1}(i+1)]$, and
$g_{l-1}[i]=g[2^{l-1}i,2^{l-1}(i+1)]$ \cite{Takahashiadd2}. The
CARRY$_l$ gate uses $\sum_{t=l}^{\lfloor \log n \rfloor -1}(\lfloor
n/2^t\rfloor -1)$ ancillary qubits and its depth and size are $O(\log
n-l)$ and $O(\sum_{t=l}^{\lfloor \log n \rfloor -1}(\lfloor n/2^t\rfloor
-1)),$ respectively. Draper et al.'s quantum carry-lookahead adder uses
$O(n)$ ancillary qubits and its depth and size are $O(\log n)$ and
$O(n)$, respectively.

In Takahashi et al.'s combination method, the input binary number
$a_{n-1}\cdots a_0$ is divided into $n/k$ blocks of length $k$, where we
assume that $n$ is a power of two for simplicity and set $k=2^{\lfloor
\log\log n\rfloor}$ and $l=\lfloor \log\log n\rfloor +1$. Note that $k =
\Theta(\log n)$ and $n$ is divisible by $k$. That is, we consider a
$k$-bit binary number $a(j)=a_{(j+1)k-1}\cdots a_{jk}$ for $0\leq j \leq
n/k-1$. Similarly, we consider $b(j)$ for $b_{n-1}\cdots b_0$. Roughly
speaking, the previous method is described as follows:
\begin{enumerate}
\item Compute the high-order bit of $a(j)+b(j)$, which is
      $g_{l-1}[j]=g[jk,(j+1)k]$, using the ripple-carry approach 
      \cite{Takahashiaddition} for $0\leq j \leq n/k-1$.

\item Compute the value $\bigwedge_{i=0}^{k-1}(a_{jk + i}\oplus b_{jk + i}),$
      which is $p_{l-1}[j]=p[jk,(j+1)k]$, using Barenco et al.'s circuit
      for a generalized Toffoli operation $T_k$ \cite{Barenco} for
      $0\leq j \leq n/k-1$, where $T_t$ (on $t+1$ qubits) is defined as
      $$T_t\left(|y\rangle\bigotimes_{i=0}^{t-1}|x_i\rangle\right)
      =|y\oplus \bigwedge_{i=0}^{t-1} x_i
      \rangle\bigotimes_{i=0}^{t-1}|x_i\rangle.$$

\item Compute the carry bit $c_{jk}=g[0,jk]$ using the values computed
      in Steps 1 and 2 for $1\leq j \leq n/k$. This is done by using the
      CARRY$_l$ gate.

\item Compute the carry bit $g[0,i]$ using the carry bits computed in
      Step 3 for $1 \leq i \leq n$ and obtain $s_i$ for $0 \leq i \leq
      n$. This is done by a circuit based on the ripple-carry approach
      as in Step 1.
\end{enumerate}
The whole circuit uses $O(n/k)$ $(=O(n/\log n)$) ancillary qubits and
its depth and size are $O(k)$ $(=O(\log n))$ and $O(n)$, respectively.

\subsection{Our Method}

Our idea is to divide the input binary numbers into $n/d(n)$ blocks of
length $d(n)$ in Takahashi et al.'s method, where $d(n)=\Omega(\log
n)$. By using the CARRY$_{\log d(n)+1}$ gate, we can construct an
$O(d(n))$-depth $O(n)$-size circuit with $O(n/d(n))$ ancillary
qubits. This is a simple generalization of the previous method. Though
this allows us to construct an $O(d(n))$-depth circuit for any
$d(n)=\Omega(\log n)$ in contrast to the previous method, it, of course,
does not improve the previous $O(\log n)$-depth circuit.

To obtain an efficient circuit, we simplify Steps 1, 2, and 4 in the
previous method using the circuit for addition in Section 2. The
simplification of Step 4 is due to a direct application of the circuit
for addition. To simplify Steps 1 and 2, we use only the first halves of
our circuit for addition and Barenco et al.'s circuit for $T_n$
\cite{Barenco}. The first half of the circuit for addition outputs the
high-order bit of $a(j)+b(j)$ and appropriate inputs to Barenco et al.'s
circuit. We use only the first half and we can thus save Toffoli gates,
but some qubits represent unuseful values. An important point is that
Barenco et al.'s circuit can use these qubits as uninitialized ancillary
qubits. We use the first half of Barenco et al.'s circuit and we can
thus again save Toffoli gates, but some qubits have unuseful
values. This is not a problem since these qubits are reset to the
initial values in later steps. The details are described below.

\begin{figure}
\vspace{-1.4cm}
\centering
\epsfig{file=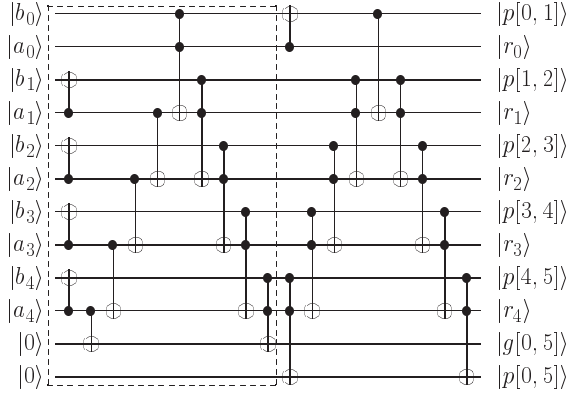,scale=.37}
\vspace{-.7cm}
\caption{The INIT$_5$ gate. A dashed-line box represents the part for
 computing $g[0,5]$, which is the first half of our circuit for addition
 in Section 2.}
\label{highbit}
\end{figure}

To simplify Steps 1 and 2, since we need to compute only the two bits
$g[i,j]$ and $p[i,j]$ for some $i,j$, it suffices to construct an
efficient quantum circuit for the operation
$$\left(\bigotimes_{i=0}^{w-1} |b_i\rangle |a_i\rangle \right)
|0\rangle|0\rangle
\to \left(\bigotimes_{i=0}^{w-1} |p[i,i+1]\rangle |r_i\rangle\right)
|g[0,w]\rangle|p[0,w]\rangle,$$
where $a_{w-1}\cdots a_0$ and $b_{w-1}\cdots b_0$ are the input binary
numbers, $r_0=a_0$, and $r_i=a_i\oplus g[0,i]\oplus p[0,i]$ $(1 \leq i
\leq w-1)$. Let $A_i$ and $B_i$ denote the memory locations initially
storing $a_i$ and $b_i$, respectively. Let $G$ and $P$ be the memory
locations initially storing 0. Location $A_i$ will store $r_i$, $B_i$
will store $p[i,i+1]$, $G$ will store $g[0,w]$, and $P$ will store
$p[0,w]$ at the end of the computation. The circuit is defined as
follows:
\begin{enumerate}
\item Apply the first half of the circuit (for two $w$-bit binary
      numbers) in Section 2 to a tuple of memory locations $A_i$ $(0
      \leq i \leq w-1)$ and $B_i$ $(0 \leq i \leq w-1)$ and $G$.

\item Apply a CNOT gate to a pair of memory locations $A_0$ and $B_0$,
      where $A_0$ is used for the control bit.

\item Apply the first half of Barenco et al.'s circuit for $T_w$ to a
      tuple of memory locations $A_i$ $(0 \leq i \leq w-1)$ and $B_i$
      $(0 \leq i \leq w-1)$ and $P$, where $A_i$ is used as an
      uninitialized ancillary memory location.
\end{enumerate}
Step 1 writes the value $g[0,w]$ into the memory location $G$. The
memory location $A_i$ stores the value $r_i$. Step 2 writes $p[0,1]$
into the memory location $B_0$. Step 3 uses the memory location $A_i$ as
an uninitialized ancillary memory location and writes the value $p[0,w]$
into the memory location $P$. The whole circuit uses no ancillary qubits
and its depth and size are $O(w)$. We call the circuit the INIT$_w$
gate. The INIT$_5$ gate is depicted in Fig. \ref{highbit}.

\begin{figure}
\vspace{-1.4cm}
\centering
\epsfig{file=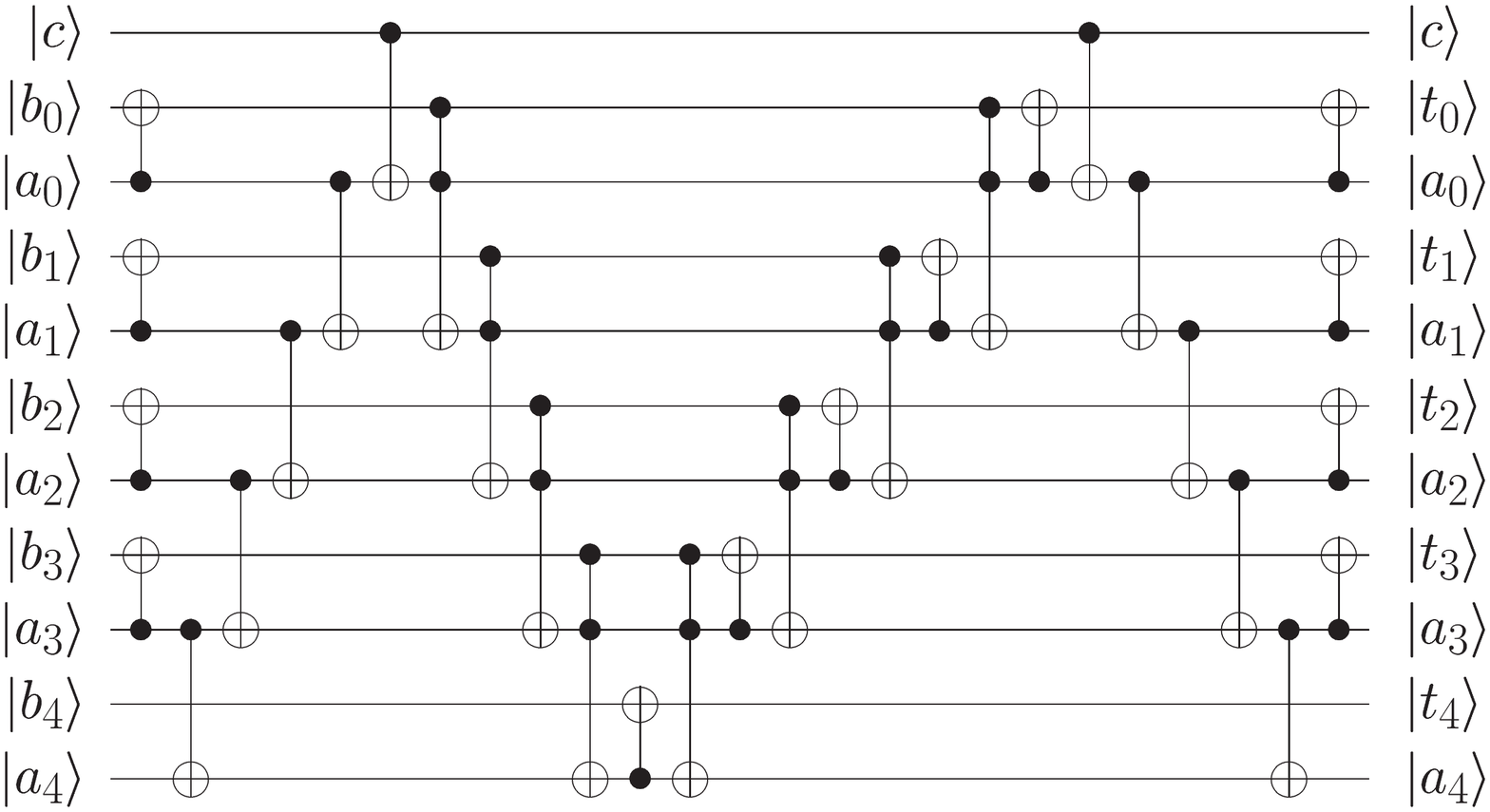,scale=.37}
\vspace{-1.2cm}
\caption{The SUM$_5$ gate.}
\label{additionincoming}
\end{figure}

To simplify Step 4, it suffices to construct an efficient quantum
circuit for the operation
$$\left(|c\rangle\bigotimes_{i=0}^{w-1} |b_i\rangle |a_i\rangle \right)
\to \left(|c\rangle\bigotimes_{i=0}^{w-1} |t_i\rangle
|a_i\rangle\right),$$
where $c\in \{0,1\}$, $a_{w-1}\cdots a_0$ and $b_{w-1}\cdots b_0$ are
the input binary numbers, $t_j=a_j\oplus b_j \oplus d_j$ $(0 \leq j \leq
w-1)$, and $d_j$ is defined as
$$ d_j = \left\{
\begin{array}{ll}
c & \mbox{$j=0$,}\\
{\rm MAJ}(a_{j-1},b_{j-1},d_{j-1}) & \mbox{$1\leq j \leq w-1$.}
\end{array}
\right.$$
We can directly apply the circuit in Section 2 to constructing such a
circuit and thus omit the details. The circuit uses no ancillary qubits
and its depth and size are $O(w)$. We call the circuit the SUM$_w$
gate. The SUM$_5$ gate is depicted in Fig. \ref{additionincoming}.

\subsection{The Whole Circuit}

We construct a quantum circuit for ADD$_n$. For simplicity, we assume
that $n$ is a power of two. Let $d(n)=\Omega(\log n)$. We set
$k=2^{\lfloor \log d(n)\rfloor}$ and $l=\lfloor \log d(n)\rfloor+1$. Note
that $k=\Theta(d(n))$ and $n$ is divisible by $k$. As described in
Section 3.1, we consider $k$-bit binary numbers $a(j)$ and $b(j)$. Let
$A_i$ and $B_i$ denote the memory locations initially storing $a_i$ and
$b_i$, respectively. Let $Z$ be the memory location initially storing
$z\in\{0,1\}$. Location $A_i$ will store $a_i$, $B_i$ will store $s_i$,
and $Z$ will store $z\oplus s_n$ at the end of the computation. We
assume that there are ancillary memory locations initially storing
0. The first half of our circuit is defined as follows:
\begin{enumerate}
\item Apply the INIT$_k$ gate to memory locations storing $a(j)$ and
      $b(j)$ and to two ancillary memory locations storing 0 for $0 \leq
      j \leq n/k-1$. The gate writes $g_{l-1}[j]$ and $p_{l-1}[j]$ into
      the ancillary memory locations.

\item Apply the CARRY$_l$ gate to memory locations storing all
      $g_{l-1}[j]$ and all $p_{l-1}[j]$ and to ancillary memory
      locations storing 0. The gate writes $c_{(j+1)k}$ into the memory
      location storing $g_{l-1}[j]$ for $0 \leq j \leq n/k-1$.

\item Apply the gates in Step 1 in reverse order, where we exclude the
      gates applied to memory locations storing $c_{(j+1)k}$ for $0\leq
      j \leq n/k-1$ since we do not erase the value.

\item Apply the SUM$_k$ gate to memory locations storing $a(j+1)$ and
      $b(j+1)$ and to a memory location storing $c_{k(j+1)}$ to obtain
      $s_{k(j+1)},\ldots,s_{k(j+2)-1}$ for $0\leq j \leq n/k-2$. Apply a
      simplified gate of the SUM$_k$ gate to memory locations storing
      $a(0)$ and $b(0)$ to obtain $s_0,\ldots,s_{k-1}$.
\end{enumerate}

The last half part deletes unnecessary carry bits using the fact that
the carry bits generated for computing $a+s'$ is the same as those for
computing $a+b$, where $s'$ is the bitwise complement of $s$
\cite{Svore}.

\begin{figure}
\vspace{-.6cm}
\centering
\epsfig{file=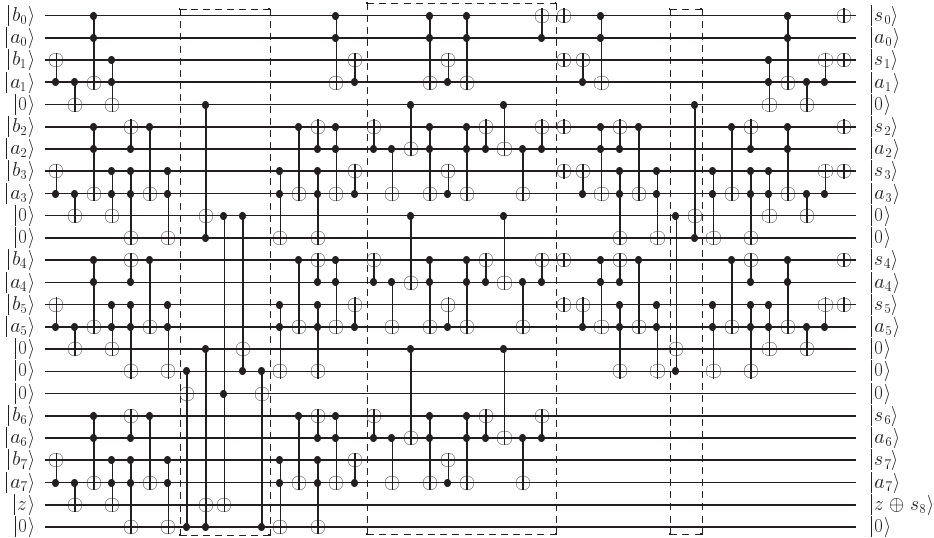,scale=.45}
\vspace{-1.2cm}
\caption{The circuit for ADD$_8$, where $d(n)=\log n$. The first and
 third dashed-line boxes represent the carry-lookahead part
 \cite{Svore,Takahashiadd2}. The second one represents the parallel
 applications of the SUM$_2$ gate.}
\label{new}
\end{figure}

\begin{enumerate}
\item[5.] Apply a NOT gate to $B_i$ to write $s_i\oplus 1$ into $B_i$
	  for $0\leq i \leq n-k-1$.

\item[6.] Apply the first half of our circuit excluding Step 4 in
	  reverse order, where we exclude the gates applied to memory
	  locations storing $a(n/k-1)$ and $b(n/k-1)$ since we do not
	  erase the last carry bit. The gate writes 0 into a memory
	  location storing $c_{k(j+1)}$ for $0\leq j \leq n/k-1$.

\item[7.] Apply a NOT gate to $B_i$ to write $s_i$ into $B_i$ for $0\leq
	  i \leq n-k-1$.
\end{enumerate}
The whole circuit for $d(n)=\log n$ and $n=8$ (and thus $k=l=2$) is
depicted in Fig. \ref{new}.

We compute the number of ancillary qubits, the depth, and the size
precisely. For simplicity, we count only Toffoli gates as in
\cite{Svore,Takahashiadd2}. Step 1 requires $\frac{2n}{k}$ ancillary
qubits to use $\frac{n}{k}$ INIT$_k$ gates. The gate consists of $3n-2$
Toffoli gates for $n\geq 3$. Thus, the depth and size of Step 1 are
$3k-O(1)$ and $3n-O(n/k)$, respectively. The CARRY$_l$ gate in Step 2
uses $\frac{n}{k}-O(\log n)$ ancillary qubits and its depth and size are
$2\log\frac{n}{k}+O(1)$ and $\frac{4n}{k}+O(\log n)$, respectively,
where $\frac{n}{k} \geq 4$ \cite{Takahashiadd2}. Step 3 is the same as
Step 1. Step 4 uses $\frac{n}{k}$ SUM$_k$ gates. The gate consists of
$2n-2$ Toffoli gates for $n\geq 3$. Thus, the depth and size of Step 4
are $2k-O(1)$ and $2n-O(n/k)$, respectively. The other steps are the
same as the above steps excluding Step 4. Our circuit uses
$\frac{3n}{k}-O(\log n)$ ancillary qubits and its depth and size are
$14k + 4\log \frac{n}{k}+O(1)$ and $14n-O(n/k)$, respectively, where
$\frac{n}{k}\geq 4$. Thus, the circuit uses $O(n/d(n))$ ancillary qubits
and its depth and size are $O(d(n))$ and $O(n)$, respectively. For
example, for $d(n)=\log n$ and $n\geq 16$, the number of ancillary
qubits, the depth, and the size are approximately $3n/\log n$, $18\log
n$, and $14n$, respectively. The corresponding previous bounds are
$3n/\log n$, $30\log n$, and $29n$. That is, in this case, the number of
ancillary qubits in our circuit is the same as that in Takahashi et
al.'s \cite{Takahashiadd2} and the leading coefficient of the expression
of the size in our circuit is less than half that in Takahashi et
al.'s.

\section{Circuit with Depth $o(\log n)$}

\subsection{Chandra et al.'s Classical Circuit}

If we use only one-qubit and two-qubit gates as elementary gates, we
cannot construct an $o(\log n)$-depth circuit for ADD$_n$. This is
simply shown by using the logarithmic lower bound for the depth of the
circuit for $F_n$ \cite{Fang}. To construct an $o(\log n)$-depth
circuit, we decrease the depth of the carry-lookahead part of our method
in Section 3 by using a quantum version of Chandra et al.'s efficient
classical circuit for addition with (classical) unbounded fan-out gates
\cite{Chandra}. We assume that we have unbounded fan-out gates
(described in Section 2) as elementary gates. We first consider the
simple case where we have unbounded fan-out gates with a long length and
then reduce the length.

Chandra et al.'s method for constructing the circuit is a generalization
of the carry-lookahead approach. Besides the (classical) unbounded
fan-out gates, the circuit uses unbounded fan-in gates that compute
logical AND (or OR) of an unbounded number of input bits. The depth and
size of the circuit for two $m$-bit binary numbers are $O(1)$ and
$O(m\log^{**}m)$, respectively, where
$$\log^{**}t =\min\{j|\overbrace{\log^*\cdots\log^*}^j t \leq 1\},\
\log^* t =\min\{j|\overbrace{\log\cdots\log}^j t \leq 1\}.$$
It can be shown that $\log^{**}m=o(\log^* m)$. Though the definition of
the depth of a classical circuit is similar to that of a quantum
circuit, the definition of the size of a classical circuit in
\cite{Chandra} is different from that of a quantum circuit. More
precisely, a classical circuit is defined as a directed acyclic graph
and the size is the number of edges in the circuit and the depth is the
length of a longest path from an input node to an output node. Chandra
et al.\ give a tighter bound on the size of the circuit, but we use the
above bound since it is sufficient for showing that our circuits in
Sections 4.2 and 4.3 use a sublinear number of ancillary qubits.

\subsection{Simple Case}

\subsubsection{Quantum Version of Chandra et al.'s Circuit}

We transform Chandra et al.'s classical circuit for two $m$-bit binary
numbers into its quantum version. Since the size (that is, the number of
edges) of the circuit is $O(m\log^{**}m)$, it suffices to consider an
unbounded fan-out gate with length $O(m\log^{**}m)$ and a $T_t$ gate
(corresponding to an unbounded fan-in gate with $t$ inputs in the
classical circuit) with $t=O(m\log^{**}m)$. We assume that we have
unbounded fan-out gates with length $O(m\log^{**}m)$. If we have
one-qubit gates, CNOT gates, $T_t$ gates, and unbounded fan-out gates
with length $O(m\log^{**}m)$, Chandra et al.'s classical circuit can be
simply transformed into its quantum version. Note that an OR gate in
Chandra et al.'s circuit is transformed into a $T_t$ gate with NOT
gates. However, in our setting, we have only one-qubit gates, CNOT
gates, and unbounded fan-out gates with length $O(m\log^{**}m)$. Thus,
we require a quantum circuit for $T_t$ (consisting of one-qubit gates,
CNOT gates, and unbounded fan-out gates with length
$O(m\log^{**}m)$). We use H\o yer et al.'s circuit for the $T_t$
operation (defined in Section 3.1) as the $T_t$ gate \cite{Hoyer}. They
showed that, if unbounded fan-out gates with length $O(t)$ are
available, an $O(\log^* t)$-depth $O(t)$-size quantum circuit for $T_t$
can be constructed. We can show that H\o yer et al.'s circuit uses
$O(t)$ ancillary qubits. Since we have unbounded fan-out gates with
length $O(m\log^{**}m)$, we can directly use H\o yer et al.'s circuit
for $T_t$ with $t=O(m\log^{**}m)$. Thus, we obtain a quantum version of
Chandra et al.'s circuit. We call the circuit the GCLA$_m$ circuit,
which stands for the generalized carry-lookahead approach for two
$m$-bit binary numbers.

The complexity of the GCLA$_m$ circuit is analyzed as follows. To
compute the depth of the circuit, since the depth of the original
circuit is $O(1)$, it suffices to consider a $T_{t_1}$ gate, where $t_1$
is the maximum number of inputs of $T_t$ gates in the GCLA$_m$
circuit. The depth of the $T_{t_1}$ gate is $O(\log^* t_1)$. Since $t_1
= O(m\log^{**}m)$, the depth of the $T_{t_1}$ gate is $O(\log^*
(m\log^{**}m))$ and thus the depth of the GCLA$_m$ circuit is $O(\log^*
(m\log^{**}m))$. To compute the size of the circuit, we define $A_t$ as
the number of unbounded fan-in gates with $t$ inputs in Chandra et al.'s
circuit, which is equal to the number of $T_t$ gates in the GCLA$_m$
circuit. Since the size of Chandra et al.'s circuit is $O(m\log^{**}m)$,
$\sum_ttA_t=O(m\log^{**}m)$. The size of a $T_t$ gate is $O(t)$. The
number of the other gates in the GCLA$_m$ circuit is $O(m\log^{**}m)$
(and the size of each gate is 1). Thus, the size of the GCLA$_m$ circuit
is $O(\sum_ttA_t+m\log^{**}m)=O(m\log^{**}m)$. A similar argument shows
that the number of ancillary qubits in the GCLA$_m$ circuit is
$O(m\log^{**}m)$. That is, the GCLA$_m$ circuit uses $O(m\log^{**}m)$
ancillary qubits and its depth and size are $O(\log^* (m\log^{**}m))$
and $O(m\log^{**}m)$, respectively.

\subsubsection{Modification of Our Method}

We modify our method in Section 3.3 by using the GCLA$_m$ circuit as the
CARRY$_l$ gate. Let $e(n) = \Omega(\log^* n)$. We set $k$ and $l$ as in
Section 3.3. Note that $k=2^{l-1}=\Theta(e(n))$. We assume that we are
allowed to use unbounded fan-out gates with length $O(n)$. Chandra et
al.'s circuit for two $\lfloor n/2^{l-1}\rfloor$-bit binary numbers is
directly applied to perform the operation performed by the CARRY$_l$
gate. Thus, we set $m=\lfloor n/2^{l-1}\rfloor$. In this case,
$O(m\log^{**}m)=O(n\log^{**}(n/2^{l-1})/2^{l-1})$, which is bounded by
$O(n)$. Since we have unbounded fan-out gates with length $O(n)$, we can
use the complexity analysis described in Section 4.2.1. The GCLA$_m$
circuit, which is the CARRY$_l$ gate, uses
$O(n\log^{**}(n/2^{l-1})/2^{l-1})$ ancillary qubits and its depth and
size are $O(\log^* (n\log^{**}(n/2^{l-1})/2^{l-1}))$ and
$O(n\log^{**}(n/2^{l-1})/2^{l-1})$, respectively. For simplicity, we
consider slightly weaker bounds; it uses $O(n\log^{**}n/2^{l-1})$
ancillary qubits and its depth and size are $O(\log^*
(n\log^{**}n/2^{l-1}))$ and $O(n\log^{**}n/2^{l-1})$, respectively.

The complexity of the whole circuit obtained by the modified method is
analyzed as in the original method. Step 1 uses $O(n/k)$ ancillary
qubits and its depth and size are $O(k)$ and $O(n)$, respectively. Step
2 uses $O(n\log^{**}n/k)$ ancillary qubits and its depth and size are
$O(\log^* (n\log^{**}n/k))$ and $O(n\log^{**}n/k)$, respectively. Step 4
requires no new ancillary qubits and its depth and size are $O(k)$ and
$O(n)$, respectively. The other steps are similar to the above
steps. Thus, the whole circuit uses $O(n\log^{**}n/e(n))$ $(=o(n))$
ancillary qubits and its depth and size are $O(e(n))$ and $O(n)$,
respectively. In particular, for $e(n)=\log^* n$, the modified method
yields an $O(\log^* n)$-depth $O(n)$-size circuit with
$O(n\log^{**}n/\log^* n)$ $(=o(n))$ ancillary qubits.

\subsection{Reduction of the Length of an Unbounded Fan-Out Gate}

We prove that the length of an unbounded fan-out gate can be restricted
to $O(n^{\varepsilon})$ in the modified method without increasing the
complexity of the circuit, where $\varepsilon$ is any small positive
constant. Suppose that we are allowed to use unbounded fan-out gates
with length $f(n)$. An unbounded fan-out gate with length
$t=O(m\log^{**}m)$ (and $m=\lfloor n/2^{l-1}\rfloor$) can be simply
simulated by using an $O(\log t/\log f(n)+1)$-depth $O(t/f(n)+1)$-size
circuit with no ancillary qubits that consists only of unbounded fan-out
gates with length $f(n)$. In the following, using this simulation, we
reconsider the complexity of the $T_t$ gate, the GCLA$_m$ circuit, and
the circuit our method in Section 4.2 yields.

\subsubsection{$T_t$ gate}

The $T_t$ gate, which is H\o yer et al.'s circuit for the $T_t$
operation, is constructed as follows:
\begin{enumerate}
\item Construct an $O(1)$-depth $O(t\log t)$-size circuit with $O(t\log
      t)$ ancillary qubits for reducing the computation of OR of $t$
      bits to that of $O(\log t)$ bits.

\item Using the circuit in Step 1, for any $d>0$, construct an
      $O(d+\log^* t)$-depth $O(dt\log^{(d)}t)$-size circuit for $T_t$
      with $O(dt\log^{(d)}t)$ ancillary qubits, where $\log^{(d)}t$ is
      the $d$-times iterated logarithm $\log\cdots\log t$.

\item Using the circuit in Step 2, construct an $O(\log^* t)$-depth
      $O(t)$-size circuit for $T_t$ with $O(t)$ ancillary qubits.
\end{enumerate}
We can modify the above steps using unbounded fan-out gates with length
$f(n)$ as follows:
\begin{enumerate}
\item Construct an $O(\log t/\log f(n)+1)$-depth $O(t\log t)$-size
      circuit with $O(t\log t)$ ancillary qubits for reducing the
      computation of OR of $t$ bits to that of $O(\log t)$ bits.

\item Using the circuit in Step 1, for any $d>0$, construct an
      $O(d+\log^* t+\log t/\log f(n)+d\log\log t/\log f(n))$-depth
      $O(dt\log^{(d)}t)$-size circuit for $T_t$ with $O(dt\log^{(d)}t)$
      ancillary qubits.

\item Using the circuit in Step 2, construct an $O(\log t/\log
      f(n)+\log^* t)$-depth $O(t)$-size circuit for $T_t$ with $O(t)$
      ancillary qubits.
\end{enumerate}
To see this, we first analyze Step 1 in H\o yer et al.'s
construction. In this step, an unbounded fan-out gate with length
$O(\log t)$ is used in parallel to make $O(\log t)$ copies of each of
the $t$ input bits. Moreover, an unbounded fan-out gate with length
$O(t)$ is used in parallel to prepare appropriate ancillary qubits
$O(\log t)$ times. As described above, an unbounded fan-out gate with
length $O(\log t)$ can be simulated by using an $O(\log\log t/\log
f(n)+1)$-depth $O(\log t/f(n)+1)$-size circuit with no ancillary
qubits. Similarly, an unbounded fan-out gate with length $O(t)$ can be
simulated by using an $O(\log t/\log f(n)+1)$-depth $O(t/f(n)+1)$-size
circuit. Thus, the depth of the $T_t$ gate is $O(\log  t/\log
f(n)+1)$. The size is $O(t \cdot(\log t/f(n) + 1) + (\log t)\cdot
(t/f(n)+1)) = O(t\log t)$. These simulations do not require any
ancillary qubits. That is, in Step 1, the number of ancillary qubits and
size remain unchanged even if we consider unbounded fan-out gates with
length $f(n)$. Thus, they also do so in Steps 2 and 3. Step 2 of H\o yer
et al.'s construction is done by using Step 1 $O(\log^* t)$ times to
reduce the computation of OR of $t$ bits to that of a constant number of
bits. Step 3 is done by reducing the computation of OR of $t$ bits to
that of $t/\log^* t$ bits and by using Step 2 with $d=\log ^* t$. These
procedures can be simply applied to the case where we use unbounded
fan-out gates with length $f(n)$ and imply the desired depth bound.

\subsubsection{The GCLA$_m$ circuit}

To compute the depth of the GCLA$_m$ circuit, it suffices to consider a
$T_{t_1}$ gate for some $t_1$ and an unbounded fan-out gate with some
length $t_2$. The depth of the $T_{t_1}$ gate is $O(\log t_1/\log f(n)+\log^*
t_1)$ and the depth of an unbounded fan-out gate with length $t_2$ is
$O(\log t_2/\log f(n)+1)$. Since $t_1$ and $t_2$ cannot be greater than
the size of Chandra et al.'s circuit, the depth of the GCLA$_m$ circuit
is $O(\log m/\log f(n) +\log^* (m\log^{**}m))$. To compute the size, we
define $B_t$ as the number of unbounded fan-out gates with length $t$
used (implicitly) in Chandra et al.'s original circuit, which is equal
to the number of unbounded fan-out gates with length $t$ (that are not
used in $T_s$ gates for any $s$) in the GCLA$_m$ circuit. Since the size
of Chandra et al.'s circuit is $O(m\log^{**}m)$, $\sum_ttB_t =
O(m\log^{**}m)$. If $t \geq f(n)$, an unbounded fan-out gate with length
$t$ can be simulated by an $O(t/f(n))$-size circuit. Thus, the size
related to unbounded fan-out gates with length greater than or equal to
$f(n)$ in the GCLA$_m$ circuit (that is, $\sum_{t\geq f(n)}(t/f(n))B_t$)
is $O(m\log^{**}m)$ since $\sum_ttB_t = O(m\log^{**}m)$. The size
related to the $T_t$ gates (that is, $O(\sum_ttA_t)$) is
$O(m\log^{**}m)$. The number of the other gates is $O(m\log^{**}m)$ (and
the size of each gate is 1). Thus, the size of the GCLA$_m$ circuit is
$O(m\log^{**}m)$. The number of ancillary qubits is the same as the
size. That is, the GCLA$_m$ circuit uses $O(m\log^{**}m)$ ancillary
qubits and its depth and size are $O(\log m/\log f(n) +\log^*
(m\log^{**}m))$ and $O(m\log^{**}m)$, respectively. Since $m=\lfloor
n/2^{l-1}\rfloor$, the circuit uses $O(n\log^{**}(n/2^{l-1})/2^{l-1})$
ancillary qubits and its depth and size are $O(\log(n/2^{l-1})/\log f(n)
+ \log^* (n\log^{**}(n/2^{l-1})/2^{l-1}))$ and
$O(n\log^{**}(n/2^{l-1})/2^{l-1})$, respectively. For simplicity, we
consider slightly weaker bounds; it uses $O(n\log^{**}n/2^{l-1})$
ancillary qubits and its depth and size are $O(\log n/\log f(n) + \log^*
(n\log^{**}n/2^{l-1}))$ and $O(n\log^{**}n/2^{l-1})$, respectively.

\subsubsection{Our Circuit}

We set $f(n)=n^{\varepsilon}$ and use the GCLA$_m$ circuit as the
CARRY$_l$ gate, where $\varepsilon$ is any small positive constant. In
this case, the CARRY$_l$ gate uses $O(n\log^{**}n/2^{l-1})$ ancillary
qubits and its depth and size are $O(\log^* (n\log^{**}n/2^{l-1}))$ and
$O(n\log^{**}n/2^{l-1})$, respectively. This is the same situation as
that in Section 4.2 except that the length of an unbounded fan-out gate
in the CARRY$_l$ gate is at most $n^{\varepsilon}$. Thus, the whole
circuit uses $O(n\log^{**}n/e(n))$ $(=o(n))$ ancillary qubits and its
depth and size are $O(e(n))$ and $O(n)$, respectively. If we set
$e(n)=\log^* n$, we obtain an $O(\log^* n)$-depth $O(n)$-size circuit
with $o(n)$ ancillary qubits.

It is worth noting that the above method for constructing a circuit for
ADD$_n$ yields an $o(\log n)$-depth $O(n)$-size circuit with $o(n)$
ancillary qubits using unbounded fan-out gates with a small length. For
example, we set $f(n)=\log n$ and $d(n)=\log n/\log\log n$. In this
case, the CARRY$_l$ gate uses $O(n\log^{**}n\log\log n/\log n)$
ancillary qubits and its depth and size are $O(\log n/\log\log n)$ and
$O(n\log^{**}n\log\log n/\log n)$, respectively. This yields an $O(\log
n/\log\log n)$-depth $O(n)$-size circuit with $O(n\log^{**}n\log\log
n/\log n)$ ancillary qubits. Such an $o(\log n)$-depth circuit cannot be
constructed by using a quantum circuit only with gates on a bounded
number of qubits \cite{Fang} or by using a classical circuit only with
bounded fan-in and unbounded fan-out gates \cite{Pippenger}. Hence,
unbounded fan-out gates even with a small length are useful for
constructing efficient quantum circuits for addition.

\section{Application}

We consider the prime field \gfp \ for some prime $p>3$. An elliptic
curve $E$ over \gfp \ is the set of points $(x,y)\in$ \gfp \ $\times$
\gfp \ satisfying $y^2 = x^3 + ax + b$, where the constants $a,b \in$
\gfp \ and $4a^3 + 27 b^2 \neq 0$, together with the point at infinity
$\cal O$. It is known that the addition operation in $E$ can be defined
and that $E$ with the addition operation forms an abelian group with
$\cal O$ serving as its identity \cite{Hankerson}. Let $P \in E$,
$\langle P \rangle$ be the subgroup of $E$ generated by $P$, and
$|\langle P \rangle|$ be the order of $\langle P \rangle$. The discrete
logarithm problem over the elliptic curve $E$ with respect to the base
$P$ is defined as follows: Given a point $Q \in \langle P \rangle$, find
the integer $0 \leq d \leq |\langle P \rangle|-1$ such that
$Q=dP$. Shor's discrete logarithm algorithm solves the problem in time
polynomial in the length of the binary representation for $|\langle P
\rangle|$ with high probability \cite{Shor}. As in \cite{Proos}, we
assume that the length of the binary representation for $|\langle P
\rangle|$ is equal to that of the binary representation for $p$.

Proos et al.\ constructed an efficient quantum circuit for Shor's
discrete logarithm algorithm for elliptic curves over \gfp \
\cite{Proos}. Let $n$ be the length of the binary representation for
$p$. The depth and size  of the circuit are $O(n^3)$. The dominant cost
is $O(n^2)$ applications of an $O(n)$-depth $O(n)$-size quantum circuit
for ADD$_n$ with $n$ ancillary qubits. For counting the number of qubits
in the circuit, it suffices to count the number of qubits in the circuit
for division in \gfp \ that maps $|x\rangle|y\rangle$ to
$|x\rangle|y/x\rangle$ for $x \ (\neq 0),y \in$ \gfp. The circuit for
division in \gfp \ uses about $5n$ qubits: $2n$ qubits are used for the
input register and about $3n$ qubits are used in the circuit for the
extended Euclidean algorithm. In the circuit for the extended Euclidean
algorithm, about $2n$ qubits are used for the input binary numbers and
intermediate results, and $n$ qubits are used for ancillary qubits
during ADD$_n$.

By simply replacing Proos et al.'s circuit for ADD$_n$ with our circuit
in Section 2, we can eliminate the $n$ ancillary qubits during
ADD$_n$ since our circuit for ADD$_n$ does not use any ancillary
qubits. The resulting circuit uses about $4n$ qubits. Since 
Proos et al.\ do not describe the precise depth or size of 
their circuit for ADD$_n$, we cannot compare the depth or size of the
resulting circuit with that of the original one precisely. However,
the depth and size of our circuit for ADD$_n$ are asymptotically the
same as those of Proos et al.'s. Thus, the depth and size of the
resulting circuit are asymptotically the same as those of the original
circuit.

By adding $o(n)$ ancillary qubits to the circuit obtained above, we can
decrease the depth asymptotically. As shown in Section 3, for any
$d(n)=\Omega(\log n)$, we have an $O(d(n))$-depth $O(n)$-size circuit
for ADD$_n$ with $O(n/d(n))$ ancillary qubits. If we use this circuit
as above, we obtain $O(n^2d(n))$-depth $O(n^3)$-size circuit for Shor's
discrete logarithm algorithm with $4n+O(n/d(n))$ qubits. Moreover, as
shown in Section 4, if we are allowed to use unbounded fan-out gates
with length $O(n^{\varepsilon})$ for an arbitrary small positive
constant $\varepsilon$, we have an $O(e(n))$-depth $O(n)$-size circuit
for ADD$_n$ with $o(n)$ ancillary qubits for any $e(n)=\Omega(\log^*
n)$. This circuit yields an $O(n^2e(n))$-depth $O(n^3)$-size circuit for
Shor's discrete logarithm algorithm with $4n+o(n)$ qubits. We can also
use the previous circuits for ADD$_n$ to improve Proos et al.'s
circuit. However, they do not yield more efficient quantum circuits for
Shor's discrete logarithm algorithm than our circuits described
above. This is simply because our circuits for ADD$_n$ is more efficient
than the previous ones.

\section{Conclusions and Future Work}

We constructed an $O(n)$-depth $O(n)$-size quantum circuit for ADD$_n$
with no ancillary qubits. The size is less than that of any other
quantum circuit ever constructed for ADD$_n$ with no ancillary
qubits. Using the circuit, we proposed a method for constructing a
small-size quantum circuit for ADD$_n$ with a small number of qubits
that has a given depth. In particular, we showed that, if we are allowed
to use unbounded fan-out gates with length $O(n^{\varepsilon})$ for an
arbitrary small positive constant $\varepsilon$, we can construct an
$O(\log^* n)$-depth $O(n)$-size circuit with $o(n)$ ancillary qubits. We
applied our circuits to constructing efficient quantum circuits for
Shor's discrete logarithm algorithm.

Interesting challenges would be to find ways of improving the quantum
circuits described in this paper. For example, can we construct an
$O(\log n)$-depth $O(n)$-size quantum circuit for ADD$_n$ with $O(1)$
ancillary qubits? Can we construct an $O(1)$-depth $O(n)$-size quantum
circuit for ADD$_n$ with $O(n)$ ancillary qubits using unbounded fan-out
gates? In the classical case, we cannot construct an $O(1)$-depth
$O(n)$-size (that is, the number of edges) circuit for addition with
unbounded fan-in and fan-out gates \cite{Dolev}.

\section*{Acknowledgments}

The authors thank Yasuhito Kawano and Go Kato for their helpful
comments.

\end{document}